\documentclass{article}

\usepackage{amssymb}
\begin{document}

\title{\textbf{Variation of mass in primordial nucleosynthesis as a test of
Induced Matter Brane Gravity }}
\author{S. Jalalzadeh$^1$\thanks{%
email: s-jalalzadeh@sbu.ac.ir} and A. M. Yazdani$^{2,3}$\thanks{%
email: yazdani@quantum.physics.uvt.ro} \\
%EndAName
$^1${\small Department of Physics, Shahid Beheshti University, Evin, Tehran
19839, Iran}\\
${^2}${\small Department of physics, Islamic Azad University of Qom, Qom, Iran}\\
${^3}${\small West University of timi\c{s}oara, V. P\^{a}rvan Ave. 4, RO-300223,
 Timi%
\c{s}oara, Romania }}
\maketitle

\begin{abstract}
The variation of  mass in induced matter theory using Ceroch-Stewart-Walter
perturbations of submanifolds \cite{1} is redefined. It is shown that the deviation of primordial
Helium production due to a variation on the difference between the ``rest''
mass of the nucleus is in agrement with induced matter brane gravity. \vspace{5mm}%
\end{abstract}

\section{Introduction}

The aim of purely geometrical description of all physical interactions as
well as that of a geometrical origin of matter as dreamed by Einstein \cite{2} has attracted a lot of interest. A gravitational theory in which the
matter is absorbed into the field itself, is called unified field theory.
There exists various extensions of Einstein's framework extracting matter
from pure geometry. Much of the works trying to extend our knowledge of
gravitational and other interactions has been concentrated on developing
theories in more than four dimensions, like supergravity \cite{3},
superstrings \cite{4} and various Kaluza-Klein (KK) theories \cite{5}. In
this theories the added extra dimensions are usually taken to be compact. To
solve the problem of non observability of the small ``internal'' space
spanned by the extra dimensions, it is usually assumed that the size of the
extra dimensions are  of the  order of Planck length, being itself a consequence of
dynamical evolution of the higher-dimensional universe, as a result of the
introducing  the higher-dimensional stress-energy tensor. On the other
hand, in \cite{6}, the authors show that the gravitational models with
compact extra dimensions, linearly perturbed Einstein equations are in
conflict with observation. There exist another extensions of Einstein's
theory in which our spacetime is a submanifold (Brane) embedded in a higher
dimensional manifold (Bulk). A revised KK approach in this direction in
which the higher-dimensional stress-energy tensor is taken to be identically
zero is the Wesson Induced Matter Theory (IMT) \cite{7}. The starting points are
the vacuum $5D$ Einstein gravitational field equations,
\begin{eqnarray}  \label{1}
\mathcal{R}_{AB}=0 \hspace{1 cm} (A,B = 0...4),
\end{eqnarray}
where $\mathcal{R}_{AB}$ is the Ricci scalar of the bulk space. The induced
field equations on the brane becomes \cite{8}
\begin{eqnarray}  \label{2}
G_{\mu\nu} = Q_{\mu\nu} -\varepsilon \mathcal{E}_{\mu\nu},\hspace{.5cm}%
(\mu,\nu=0,...,3),
\end{eqnarray}
where $\mathcal{E}_{\mu\nu}$ is electric part of Weyl tensor of the bulk
space and $Q_{\mu\nu}$ is defined as
\begin{eqnarray}  \label{3}
Q_{\mu\nu} =\varepsilon\left[ K_\mu^\gamma K_{\gamma\nu} - KK_{\mu\nu} -%
\frac{1}{2}\left(K_{\alpha\beta}K^{\alpha\beta} - K^2\right)g_{\mu\nu}\right]%
,
\end{eqnarray}
where $g_{\mu\nu}$ is the induced metric on the brane, $\varepsilon = \pm 1$
denotes the signature of the extra dimension, $K_{\mu\nu}$ is the extrinsic
curvature and $K$ is its trace. The reason that this theory is called
induced matter theory (IMT)  is that the effective $4D$ matter is a consequence
of the geometry of the bulk \cite{8}
\begin{eqnarray}  \label{4}
-8\pi G_NT_{\mu\nu} = Q_{\mu\nu}-\varepsilon\mathcal{E}_{\mu\nu}.
\end{eqnarray}
One of the outcomes of IMT is that the ``rest mass'' of particles varies
from point to point in spacetime, in agreement with the ideas of Mach. To
show the variation of mass, Wesson by using dimensional analysis \cite{9}
introduced the following relation between of fifth coordinate and the mass
of the test particles
\begin{eqnarray}\label{5}
x^4 = \frac{G_N m}{c^2},
\end{eqnarray}
where $m$ is the ``rest'' mass of a test particle and $x^4$ denotes the
fifth dimension. Hence, according to the above equation if we consider that
a variation of the rest mass of  particles had occurred between the
epoch of primordial nucleosynthesis and present, we can compute the deviation
in the $^4He$ production from Hot Big Bang model production due to this
fact. In \cite{10} the authors show that if we use the relation (\ref{5})
to obtain the variation of mass from primordial nucleosynthesis and our time,
and compare with variation of mass obtained from nucleosynthesis bounds on
the variation of the mass, the results are not in agreement with each other.
They used the $5D$ metric with compact extra dimension. In this paper we
reobtain the variation of mass in IMT according to the resent developments
in this theory and in a simple model it is showed that by correct defining
of the induced mass, the variation of mass obtained from IMT is in agreement
with mass variation bound obtained from Hot Big Bang.

\section{Test particle dynamics and induced mass}

In this section we wish to derive the $4D$ geodesic equations and induced
mass of a test particle. To doing this, we start with the induced parallel
displacement in $4D$. According to the recent developments in IMT, the
assumption is that that our spacetime can be isometrically and locally
embedded in a Ricci-flat $5D$ spacetime. In contrast to the Randall and
Sundrum brane models where the matter field is confined to the fixed brane,
in IMT there is no mechanism to confine induced matter field exactly on a
specific brane. The authors of \cite{11} and \cite{12} show that to
confine test particles on a brane it is necessary to exist  either a
non-gravitational centripetal confining force with an unknown source, or
assume that our brane is totaly geodesic in which case it is impossible to
embed an arbitrary brane in the bulk space. In IMT however, if the induced
matter field satisfies ``machian strong energy condition''
then the test particles become stable around the fixed brane \cite{13}.
Finally, we can say that in IMT at the large scales we have matter field
confined to a fixed brane, say $\bar{g}_{\mu \nu }$ that satisfies induced
Einstein field equations  and at small scales we find the matter fields having small fluctuations around this brane \cite{14}. If we denote the
metric of this brane by $g_{\mu \nu }$, then it becomes acceptable to assume
that this new brane is a perturbation of the original one $\bar{g}_{\mu \nu}$ \cite{15}. In the following we briefly review the relation of
geometrical objects in these two branes, for more details  see \cite{16}.%
\newline
Consider the background manifold $\overline{V}_{4}$ isometrically embedded
in ${V}_{5}$ by a map $\mathcal{Y}:\overline{V}_{4}\rightarrow V_{5}$ such
that
\begin{eqnarray}\label{6}
\mathcal{G}_{AB}\mathcal{Y}_{\,\,\,,\mu }^{A}\mathcal{Y}_{\,\,\,,\nu }^{B}=%
\bar{g}_{\mu \nu },\hspace{0.5cm}\mathcal{G}_{AB}\mathcal{Y}_{\,\,\ ,\mu
}^{A}\mathcal{N}^{B}=0,\hspace{0.5cm}\mathcal{G}_{AB}\mathcal{N}^{A}\mathcal{%
N}^{B}=\varepsilon
\end{eqnarray}
where $\mathcal{G}_{AB}$ $(\bar{g}_{\mu \nu })$ is the metric of the bulk
(brane) space $V_{5}(\overline{V}_{4})$ in an arbitrary coordinate with
signature $(-,+,+,+,\varepsilon )$, $\{\mathcal{Y}^{A}\}$ $(\{x^{\mu }\})$
are the basis of the bulk (brane) and $\mathcal{N}^{A}$ is a normal unit
vector orthogonal to the brane. Perturbation of $\overline{V}_{4}$ in a
sufficiently small neighborhood of the brane along an arbitrary transverse
direction $\zeta $ is given by 
\begin{eqnarray}\label{7}
\mathcal{Z}^{A}(x^{\mu },x^{4})=\mathcal{Y}^{A}+(\mathcal{L}_{\zeta }%
\mathcal{Y})^{A},
\end{eqnarray}%
where $\mathcal{L}$ represents the Lie derivative and $x^{4}$ is a small
parameter along $\mathcal{N}^{A}$ parameterizing the extra noncompact
dimension. By choosing $\zeta $ orthogonal to the brane we ensure gauge
independency  and have perturbations of the embedding along a
single orthogonal extra direction $\bar{\mathcal{N}}$, giving the local
coordinates of the perturbed brane as
\begin{eqnarray}\label{8}
\mathcal{Z}_{,\mu }^{A}(x^{\nu },x^{4})=\mathcal{Y}_{,\mu }^{A}+x^{4}\bar{%
\mathcal{N}}_{\,\,\,,\mu }^{A}(x^{\nu }).
\end{eqnarray}%
In a similar manner, one can find that since the vectors $\bar{\mathcal{N}}%
^{A}$ depend only on the local coordinates $x^{\mu }$, they do not propagate
along the extra dimension. The above assumptions lead to the embedding
equations of the perturbed geometry
\begin{eqnarray}\label{9}
\mathcal{G}_{\mu \nu }=\mathcal{G}_{AB}\mathcal{Z}_{\,\,\ ,\mu }^{A}\mathcal{%
Z}_{\,\,\ ,\nu }^{B},\hspace{0.5cm}\mathcal{G}_{\mu 4}=\mathcal{G}_{AB}%
\mathcal{Z}_{\,\,\ ,\mu }^{A}\mathcal{N}^{B},\hspace{0.5cm}\mathcal{G}_{AB}%
\mathcal{N}^{A}\mathcal{N}^{B}=\mathcal{G}_{44}.
\end{eqnarray}%
If we set $\mathcal{N}^{A}=\delta _{4}^{A}$, then the line element of the
bulk space in the Gaussian frame (\ref{9}) becomes
\begin{eqnarray}\label{10}
dS^{2}=\mathcal{G}_{AB}d\mathcal{Z}^{A}d\mathcal{Z}^{B}=g_{\mu \nu
}(x^{\alpha },x^{4})dx^{\mu }dx^{\nu }+\varepsilon (dx^{4})^{2},
\end{eqnarray}%
where
\begin{eqnarray}\label{11}
g_{\mu \nu }=\bar{g}_{\mu \nu }-2x^{4}\bar{K}_{\mu \nu }+(x^{4})^{2}\bar{g}%
^{\alpha \beta }\bar{K}_{\mu \alpha }\bar{K}_{\nu \beta },
\end{eqnarray}%
is the metric of the perturbed brane, so that
\begin{eqnarray}\label{12}
\bar{K}_{\mu \nu }=-\mathcal{G}_{AB}\mathcal{Y}_{\,\,\,,\mu }^{A}\mathcal{N}%
_{\,\,\ ;\nu }^{B},
\end{eqnarray}%
represents the extrinsic curvature of the original brane. Any fixed $x^{4}$
signifies a new perturbed brane, enabling us to define an extrinsic
curvature similar to the original one by
\begin{eqnarray}\label{13}
{K}_{\mu \nu }=-\mathcal{G}_{AB}\mathcal{Z}_{\,\,\ ,\mu }^{A}\mathcal{N}%
_{\,\,\ ;\nu }^{B}=\bar{K}_{\mu \nu }-x^{4}\bar{K}_{\mu \gamma }\bar{K}%
_{\,\,\ \nu }^{\gamma }.
\end{eqnarray}%
The above perturbation is needed in the reminding of the paper. To obtain
induced parallel displacement and and the mass, consider an arbitrary vector in $%
5D$ bulk space $X_{A}$ that has a $4D$ counterpart in a brane in which the vector $%
X_{\mu }$ is defined. These two vectors are related by the following inducing relation
\begin{eqnarray}\label{14}
X_{\mu }=\mathcal{G}_{AB}X^{A}\mathcal{Z}_{,\mu }^{B}.
\end{eqnarray}%
Let us consider an infinitesimal parallel displacement of a vector in the
bulk space
\begin{eqnarray}\label{15}
dX_{A}=-\bar{\Gamma}_{AC}^{B}X_{B}d\mathcal{Z}^{C},
\end{eqnarray}%
where $\bar{\Gamma}_{AC}^{B}$ denotes the Christoffel symbols of the bulk
space. Now using equation (\ref{14}) and (\ref{15}), the induced parallel
displacement of $X_{\mu }$ is
\begin{eqnarray}\label{16}
dX_{\mu }=\mathcal{G}_{AM}\bar{\Gamma}_{BC}^{M}\mathcal{Z}_{,\mu }^{B}X^{A}d%
\mathcal{Z}^{C}+\mathcal{G}_{AB}X^{A}d\mathcal{Z}_{,\mu }^{B}.
\end{eqnarray}%
As the bulk space may be mapped either by $\{\mathcal{Z}^{A}\}$ or by local
coordinates of brane and extra dimension, one can write
\begin{eqnarray}\label{17}
d\mathcal{Z}^{C}=\mathcal{Z}_{,\alpha }^{C}dx^{\alpha }+\mathcal{N}%
^{C}dx^{4}.
\end{eqnarray}%
Inserting decomposition (\ref{17}) into the expression for the parallel
displacement (\ref{16}) we obtain
\begin{eqnarray}\label{18}
dX_{\mu }=\mathcal{G}_{AM}\left( \bar{\Gamma}_{BC}^{M}\mathcal{Z}_{,\mu
}^{B}X^{A}+X^{A}\mathcal{Z}_{,\mu ,C}^{M}\right) \left\{ \mathcal{Z}%
_{,\alpha }^{C}dx^{\alpha }+\mathcal{N}^{C}dx^{4}\right\} .
\end{eqnarray}%
In The Gaussian frame (\ref{10}) this may be rewritten as
\begin{eqnarray}\label{19}
dX_{\mu }=\Gamma _{\mu \alpha }^{\beta }X_{\beta }dx^{\alpha }+K_{\mu \alpha
}X_{4}dx^{\alpha }-K_{\,\,\mu }^{\beta }X_{\beta }dx^{4},
\end{eqnarray}%
where $\Gamma _{\mu \alpha }^{\beta }$ denotes the Christoffel symbols of
the brane. In the particular case where the induced parallel
displacement  is discussed, we use  $5$-velocity vector
\begin{eqnarray}\label{20}
X^{A}=\frac{d\mathcal{Z}^{A}}{dS}=\mathcal{Z}_{,\alpha }^{A}\frac{dx^{\alpha
}}{dS}+\mathcal{N}^{A}\frac{dx^{4}}{dS},
\end{eqnarray}%
where $dS$ is line element in the bulk space. In this case $X^{A}$
represents $5$-velocity in the bulk. But it is not clear that the
parameterization of path of test particles in the bulk and brane are proportional. Accordingly, we use in general the different parameterization
on the brane. Hence, the corresponding induced component of $X^{A}$
according to the equations (\ref{14}) and (\ref{20}) becomes
\begin{eqnarray}\label{21}
X_{\mu }=\mathcal{G}_{AB}X^{A}\mathcal{Z}_{,\mu }^{B}=eg_{\alpha \mu
}u^{\alpha },
\end{eqnarray}%
where $e=\frac{d\lambda }{dS}$ and $u^{\alpha }=\frac{dx^{\alpha }}{d\lambda
}$ is $4$-velocity of test particle on the brane. Now using (\ref{19}), the
induced parallel displacement becomes
\begin{eqnarray}\label{22}
\frac{du^{\mu }}{d\lambda }+\Gamma _{\alpha \beta }^{\mu }u^{\alpha
}u^{\beta }=-\frac{\dot{e}}{e}u^{\mu }+2K_{\,\,\,\alpha }^{\mu }u^{\alpha
}u^{4},
\end{eqnarray}%
where $u^{4}=\frac{dx^{4}}{d\lambda }$ and overdot denotes derivative
respect to $\lambda $. Repeating the above process with respect to induced normal
component of $5$-velocity, we obtain
\begin{eqnarray}\label{23}
dX_{4}=\mathcal{G}_{AB}X^{A}\left( \mathcal{Z}_{,4}^{M}\bar{\Gamma}_{MC}^{B}+%
\mathcal{Z}_{,4,C}^{B}\right) \{\mathcal{Z}_{,\alpha }^{C}dx^{\alpha }+%
\mathcal{N}^{C}dx^{4}\}.
\end{eqnarray}%
In the Gaussian frame the above equation takes the following form
\begin{eqnarray}\label{24}
dX_{4}=-g_{\mu \nu }X^{\mu }K_{\alpha }^{\nu }dx^{\alpha }.
\end{eqnarray}%
Hence the equation of motion of the test particle along the normal to the brane
direction becomes
\begin{eqnarray}\label{25}
\frac{du^{4}}{d\lambda }=-K_{\mu \nu }u^{\mu }u^{\nu }-\frac{\dot{e}}{e}%
u^{4}.
\end{eqnarray}%
In the continuum let us consider the square of the length $X^{2}:=g^{\mu
\nu }X_{\mu }X_{\nu }$. Its change under parallel displacement as
\begin{eqnarray}\label{26}
dX^{2}=g_{\,\,\,,\gamma }^{\mu \nu }X_{\mu }X_{\nu }dx^{\gamma }+g_{,4}^{\mu
\nu }X_{\mu }X_{\nu }dx^{4}+2g^{\mu \nu }X_{\mu }dX_{\nu }.
\end{eqnarray}%
Making use of $g_{\,\,\,,\gamma }^{\mu \nu }=-\Gamma _{\gamma \beta }^{\mu
}g^{\nu \beta }-\Gamma _{\gamma \beta }^{\nu }g^{\mu \beta }$ and $%
g_{\,\,\,,4}^{\mu \nu }=2K^{\mu \nu }$, we obtain from equations  (\ref{19}) the change of the squared length of the $4$-vector
\begin{eqnarray}\label{27}
dX^{2}=2X^{\mu }X^{4}K_{\mu \alpha }dx^{\alpha }.
\end{eqnarray}%
Thus, in general case, the brane possesses a non-integrable geometry \cite{17},\cite{18}
, and only when the original $5D$ vectors do not have extra components, or
when the extrinsic curvature vanishes one has a pseudo-Riemannian brane. In the non-integrable geometry, there is a well known method to measure the
``length curvature" $F:=dA$, $A_\mu=K_{\mu\nu}X^\nu
X^4$ by means of the s-called ``second clock effect" . Let us assume that, we have two standard clocks which are close to each
other and synchronized in the beginning. Now if these two clocks are separated for a while and brought together again later, they will be out of synchronization
in general. This is a well known effect from general and special relativity
and called ``first clock effect" and often
called the twin paradox. The second clock effect exists if, in addition, the units of the two clocks are different after their meeting again. In
Lorentzian spacetime there is no second clock effect for standard clocks.
Assuming that atomic clocks are standard clocks, then in general, after the
above argument, they have different properties. To solve this problem Dirac
\cite{19} assumed that in practice we have two different intervals: $ds_{A}$
 and $ds_{E}$. The interval $ds_{A}$  is referred to atomic units; it is not affected by A. The Einstein interval $ds_{E}$ is associated with the field equations and the non-integrable geometry. Another solution to the problem was given by Wood and Papini \cite{20}. In their approach, the atom appears as a bubble. Outside one has the non-integrable spacetime, and on the boundary surface and in the interior of the atom we have $A_{i}=0$. The static spherical entity is filled with ``Dirac
matter" satisfying equation of state like cosmological
constant. Finally the third method is discussed by Audretsch \cite{21} and Flint \cite{22}. In this approach, the above solutions are classified as
non-quantum-mechanical ways and we can set second clock effect as a quantum
effect.

To find the induced mass on the brane, we project 5-momenta $P_{A}$ into the
brane. This projection is done by vielbeins $\mathcal{Z}_{,\mu }^{A}$, then
\begin{eqnarray}\label{28}
p_{\mu }=\mathcal{G}_{AB}P^{A}\mathcal{Z}_{,\mu }^{B}.
\end{eqnarray}%
For a $4D$ observer, the motion is described by 4-momenta (\ref{28}) such
that
\begin{eqnarray}\label{29}
g_{\mu \nu }p^{\mu }p^{\nu }=-m^{2},
\end{eqnarray}%
where $m$ is $4D$ induced mass. On the other hand, we defined 4-velocity as $%
u^{\mu }=\frac{dx^{\mu }}{d\lambda }$. Hence we have
\begin{eqnarray}\label{30}
g_{\mu \nu }u^{\mu }u^{\nu }=\left( \frac{ds}{d\lambda }\right) ^{2}\equiv
-l^{2},
\end{eqnarray}%
Now, comparing equations (\ref{29}) and (\ref{30}) we obtain
\begin{eqnarray}\label{31}
p^{\mu }=\frac{m}{l}u^{\mu }.
\end{eqnarray}%
Usually we assume that the length of 4-velocity is normalized to unity. But
in this model, the equation (\ref{21}) implies if $X^{\mu }=eu^{\mu }$ then
\begin{eqnarray}\label{32}
d\left( -e^{2}l^{2}\right) =2K_{\mu \nu }u^{\mu }u^{\nu }u^{4}d\lambda .
\end{eqnarray}%
It is well-known that in the non-integrable geometry the normal component of
acceleration vanishes\\ $u^{\mu }u^{\nu }u_{\mu ;\nu }=0$ \cite{Rosen}.
Referring to the this fact, contracting equation (\ref{22}) with 4-velocity
of the test article , the result is
\begin{eqnarray}\label{33}
\frac{\dot{e}}{e}=-\frac{2}{l^{2}}K_{\mu \nu }u^{\mu }u^{\nu }u^{4}.
\end{eqnarray}%
Inserting this result in previous equation (\ref{32}), we obtain
\begin{eqnarray}\label{34}
\frac{dl}{l}=\frac{1}{l^{2}}K_{\mu \nu }u^{\mu }u^{\nu }u^{4}d\lambda .
\end{eqnarray}%
Now we can compute the variation of the mass of test particle. Using (\ref{27})
we have
\begin{eqnarray}\label{35}
d\left( g_{\mu \nu }p^{\mu }p^{\nu }\right) =2K_{\alpha \beta }u^{\alpha
}u^{\beta }p^{4}d\lambda,
\end{eqnarray}%
or using equation (\ref{29}) and the corresponding definition of extra
momenta $p^{4}=\frac{m}{l}u^{4}$ we obtain
\begin{eqnarray}\label{36}
\frac{dm}{m}=-\frac{1}{l^{2}}K_{\mu \nu }u^{\mu }u^{\nu }u^{4}d\lambda .
\end{eqnarray}%
The author of \cite{23} obtained the same result by using
Hamilton-Jacobi formalism,  and showed that this
expression showing variation of mass is independent of the coordinates and
any parameterization used along the motion. Now we are ready to discuss the
physical meaning of the variation of mass and non-integrability. In general
relativity we deal with large scales or at least up to scales of the
order of millimeter. According to \cite{14} the influence of matter
fields on the bulk space is small and at  large scales the matter
\textquotedblleft seems\textquotedblright\ to be on the original brane $\bar{%
g}_{\mu \nu }$. For this reason, we parameterize the path of a particle with
an affine parameter in the original brane. According to (\ref{36}) and fact
that $u^{\alpha }=dx^{\alpha }/d\lambda =(dx^{\alpha }/d\tau )(d\tau
/d\lambda )$ we have
\begin{eqnarray}\label{37}
\frac{dm}{m}=-K_{\mu \nu }\bar{u}^{\mu }\bar{u}^{\nu }\bar{u}^{4}\left(
\frac{d\tau }{ds}\right) ^{2}d\tau ,
\end{eqnarray}%
where $ds^{2}=g_{\mu \nu }dx^{\mu }dx^{\nu }$ is the line element of the
perturbed brane, $d\tau ^{2}=-ds^{2}$ is the propertime defined on the original
brane and $\bar{u}^{\alpha }$ is the $4$-velocity of the test particle in
the original non-perturbed brane. Now using equation (\ref{10}) and (\ref{11}%
) we have
\begin{eqnarray}\label{38}
-\left( \frac{ds}{d\tau }\right) ^{2}=1+2x^{4}\bar{K}_{\mu \nu }\bar{u}^{\mu
}\bar{u}^{\nu }+\mathcal{O}((x^{4})^{2}),
\end{eqnarray}%
and consequently inserting equation (\ref{38}) and (\ref{13}) into equation (%\ref{37}) we obtain
\begin{eqnarray}\label{39}
\frac{dm}{m}=\left[ \frac{1}{R}-\left( \frac{2}{R^{2}}+\bar{K}_{\mu \gamma }%
\bar{K}_{\,\,\,\nu }^{\gamma }\bar{u}^{\mu }\bar{u}^{\nu }\right) x^{4}%
\right] \bar{u}^{4}d\tau ,
\end{eqnarray}%
where
\begin{eqnarray}\label{40}
\frac{1}{R}=\bar{K}_{\mu \nu }\bar{u}^{\mu }\bar{u}^{\nu }
\end{eqnarray}%
is the normal curvature \cite{24}. In fact the normal curvature is nothing
more than the higher dimensional generalization of the familiar centripetal
acceleration. Note that according to  equation (\ref{3}) the
last term in (\ref{40}) is related to the energy-momentum tensor of induced
matter. Using (\ref{3}) and (\ref{4}) one can easily show that
\begin{eqnarray}\label{41}
\bar{K}_{\mu \gamma }\bar{K}_{\,\,\,\nu }^{\gamma }\bar{u}^{\mu }\bar{u}%
^{\nu }=-8\pi G\varepsilon \left( \bar{T}_{\mu \nu }\bar{u}^{\mu }\bar{u}%
^{\nu }+\frac{1}{2}\bar{T}\right) +\frac{\bar{K}}{R}.
\end{eqnarray}%
Hence the variation of mass is given by
\begin{eqnarray}\label{42}
\frac{dm}{m}=\left[ \frac{1}{R}+\left\{ -\frac{2}{R^{2}}-\frac{\bar{K}}{R}%
+8\pi G\varepsilon \left( \bar{T}_{\mu \nu }\bar{u}^{\mu }\bar{u}^{\nu }+%
\frac{1}{2}\bar{T}\right) \right\} x^{4}\right] \bar{u}^{4}d\tau .
\end{eqnarray}%
One  thing in above the equation for computing the variation of mass is
to replace the normal component of velocity $\bar{u}^{4}$. Using approximation (%
\ref{38}), normal geodesic equation (\ref{25}) up to first order  $x^{4}$,
takes the following form
\begin{eqnarray}\label{43}
\frac{d^{2}x^{4}}{d\tau ^{2}}+\left( \frac{2}{R^{2}}+\bar{K}_{\mu \gamma }%
\bar{K}_{\,\,\,\nu }^{\gamma }\bar{u}^{\mu }\bar{u}^{\nu }\right) x^{4}-%
\frac{1}{R}=0,
\end{eqnarray}%
or
\begin{eqnarray}\label{44}
\frac{d^{2}x^{4}}{d\tau ^{2}}+\left[ \frac{2}{R^{2}}+\frac{\bar{K}}{R}-8\pi
G\varepsilon \left( \bar{T}_{\mu \nu }\bar{u}^{\mu }\bar{u}^{\nu }+\frac{1}{2%
}\bar{T}\right) \right] x^{4}-\frac{1}{R}=0.
\end{eqnarray}%
In general, there is not any general solution to the above equation. A useful
method exists for determining an approximate solutions to the above
differential equation. This is known in the literature as the
Wentzel-Kramers-Brillouim (W.K.B) method. If we set
\begin{eqnarray}\label{45}
x^{4}=A(\tau )e^{i\phi (\tau )}+B(\tau ),
\end{eqnarray}%
and substitute this solution into (\ref{44}) we obtain
\begin{eqnarray}\label{46}
\left( \frac{d^{2}A}{d\tau ^{2}}-A\left( \frac{d\phi }{d\tau }\right)
^{2}+PA\right) e^{i\phi }+i\left( 2\frac{dA}{d\tau }\frac{d\phi }{d\tau }+A%
\frac{d^{2}\phi }{d\tau ^{2}}\right) e^{i\phi }+\frac{d^{2}B}{d\tau ^{2}}%
+PB+Q=0.
\end{eqnarray}%
Therefor we obtain
\begin{eqnarray}\label{47}
\begin{array}{ll}
2\frac{dA}{d\tau }\frac{d\phi }{d\tau }+A\frac{d^{2}\phi }{d\tau ^{2}}=0, &
\\
&  \\
\frac{d^{2}A}{d\tau ^{2}}-A\left( \frac{d\phi }{d\tau }\right) ^{2}+PA=0, &
\\
&  \\
\frac{d^{2}B}{d\tau ^{2}}+PB+Q=0, &
\end{array}
\end{eqnarray}%
where
\begin{eqnarray}\label{48}
\begin{array}{cc}
P=\left[ \frac{2}{R^{2}}+\frac{\bar{K}}{R}-8\pi G\varepsilon \left( \bar{T}%
_{\mu \nu }\bar{u}^{\mu }\bar{u}^{\nu }+\frac{1}{2}\bar{T}\right) \right] ,
&  \\
&  \\
Q=-\frac{1}{R}. &
\end{array}%
\end{eqnarray}%
Since $P$ and $Q$ are assumed to vary slowly, so are $A$ and $B$, and
thus we neglect the second derivatives of $A$ and $B$. We thus obtain
\begin{eqnarray}\label{49}
x^{4}=\frac{C}{P^{\frac{1}{4}}}\exp \left( \pm i\int \sqrt{P}d\tau \right) -%
\frac{P}{Q},
\end{eqnarray}%
where $C$ is a constant of integration. This solution shows that the test
particle becomes stable around the original non perturbed brane, if $P$
becomes greater than zero. i.e.,
\begin{eqnarray}\label{50}
-8\pi G\varepsilon \left( \bar{T}_{\mu \nu }\bar{u}^{\mu }\bar{u}^{\nu }+%
\frac{1}{2}\bar{T}\right) +\frac{2}{R^{2}}+\frac{\bar{K}}{R}>0.
\end{eqnarray}%
This, in turn means that the induced energy-momentum tensor satisfies  some
kind of energy condition. As a consequence, if the energy-momentum tensor
vanishes, so does the extrinsic curvature. This means that
according to equation (\ref{44}), the particle becomes totally unstable.
Such a result seems to be in accordance with Mach's principal and for this
reason we may call the above energy condition as Machian energy condition.
Now, inserting equation (\ref{43}) into equation (\ref{42}) gives the
following result
\begin{eqnarray}\label{51}
m=m_{0}exp\left( \frac{1}{2}(\bar{u}^{4})^{2}\vert_{\bar{u}_{{in}}^{4}}^{\bar{u}%
_{fi}^{4}}\right) ,
\end{eqnarray}%
where $m_{0}$ is the initial mass, $\bar{u}_{fi}^{4}$ and $\bar{u}_{{in}}^{4}
$ denote the initial and final velocity along extra dimension respectively.
In the next section we will use this equation to obtain the variation of
mass of nucleons from Big Bang nucleosynthesis up to now.

\section{Variation of mass in FRW brane}

Consider a FRW universe embedded (as a non perturbed brane) in an $5D$ flat
bulk space so that the extra dimension is spacelike. The FRW line element is
written as
\begin{eqnarray}\label{52}
ds^{2}=-dt^{2}+a^{2}(t)\left[ \frac{dr^{2}}{1-kr^{2}}+r^{2}(d\theta ^{2}+\sin
^{2}\theta d\varphi ^{2})\right] ,
\end{eqnarray}%
where $\kappa $ takes the values $\pm 1$ or $0$ and $a(t)$ is the scale
factor. Now, we proceed to analyze the variation of mass of the test
particle. To do this, we first compute the extrinsic curvature through
solving the Codazzi equations that gives \cite{25}
\begin{eqnarray}\label{53}
\begin{array}{ll}
\bar{K}_{00}=-\frac{1}{\dot{a}}\frac{d}{dt}\left( \frac{b}{a}\right) , &  \\
&  \\
\bar{K}_{ij }=\frac{b}{a^{2}}g_{ij},\hspace{1.3cm}i,j=1,2,3. &
\end{array}
\end{eqnarray}%
Here, $b$ is an arbitrary functions of $t$. Consequently, the components of $%
\bar{Q}_{\mu \nu }$ using definition (\ref{3}) become
\begin{eqnarray}\label{54}
\begin{array}{ll}
\bar{Q}_{00}=-\frac{3}{a^{4}}b^{2}, &  \\
&  \\
\bar{Q}_{ij}=\frac{1}{a^{4}}\left( 2\frac{b\dot{b}}{H}-b^{2}\right) g_{ij}%
\hspace{0.5cm}i,j=1,2,3, &
\end{array}%
\end{eqnarray}%
where $H=\frac{\dot{a}}{a}$ is the Hubble parameter and a dote denotes
derivative with respect to the cosmological time $t$. Now the geodesic equation
along the extra dimension (\ref{44}) becomes
\begin{eqnarray}\label{55}
\frac{d^{2}x^{4}}{dt^{2}}+\frac{3}{R^{2}}x^{4}+\frac{1}{R}=0,
\end{eqnarray}%
with approximate solution
\begin{eqnarray}\label{56}
x^{4}\sim CR^{\frac{1}{2}}\sin \left( \sqrt{3}\int \frac{dt}{R}+\varphi
\right) -\frac{R}{3},
\end{eqnarray}%
so that $C$ and $\varphi $ are integration constants and
\begin{eqnarray}\label{57}
\frac{1}{R}=\bar{K}_{\mu \nu }\bar{u}^{\mu }\bar{u}^{\nu }=\bar{K}_{oo}.
\end{eqnarray}%
Since within ordinary scales of energy we do not see the disappearance of
particles, one may assume that the width of brane is very small. In
braneworld models with large extra dimension, usually the width of brane
should be in order or less than $TeV^{-1}$, i.e., $L\sim 10^{-17}cm$. On the other hand, the standard model fields are usually confined to the brane
within some localized width i.e, the brane width [38--40]. Similarly, in
Induced Matter Theory, if the induced matter satisfies the restricted energy
condition, the particles will be stabilized around the original brane [41].
The size of the fluctuations of the induced matter corresponds to the
width of the brane. Since within the ordinary scales of energy we do not see the disappearance of particles, one may assume the fluctuations of the
matter field exist only around the original brane. In other words, if the
brane width is $d$, it means that brane localized particles probe this length
scale across the brane and therefore the observer cannot measure the
distance on the brane to a better accuracy than $d$.  On the other hand, the observational data constrains the brane width to be in the order of planck length, see \cite{p1} and references therein. Hence, in this paper according to \cite{p2} we assume that the size of the fluctuations of the brane (the width of the brane) is in order of Planck length which is much smaller than the effective size of the extra dimension $L$. This assumption may help us to investigate the correct quantum phenomenology in IMT. So we can neglect effect of this term in our calculation.\newline

To proceed with geometrical interpretation of the energy-momentum tensor,
let us consider an analogy between $\bar{Q}_{\mu \nu }$ and a simple example
of matter consisting of free radiation field plus dust, that is
\begin{eqnarray}\label{58}
\bar{Q}_{\mu \nu }=-8\pi GT_{\mu \nu }+\Lambda g_{\mu \nu }=-8\pi G\left[
(p+\rho )u_{\mu }u_{\nu }+pg_{\mu \nu }\right] +\Lambda g_{\mu \nu },
\end{eqnarray}%
with equation of state
\begin{eqnarray}\label{59}
p=\omega \rho .
\end{eqnarray}%
Using equations (\ref{54}) and (\ref{58}) the energy density and pressure takes
the following forms
\begin{eqnarray}\label{60}
\begin{array}{ll}
\rho =-\frac{3}{8\pi Ga^{4}}b^{2}-\frac{1}{8\pi G}\Lambda , &  \\
\\
p=\frac{1}{8\pi Ga^{4}}\left( \frac{2b\dot{b}}{H}+b^{2}\right) -\frac{1}{%
8\pi G}\Lambda . &
\end{array}%
\end{eqnarray}%
Using the above two equations and equation of state of the matter we obtain
\begin{eqnarray}\label{61}
\begin{array}{ll}
\Lambda =\frac{b^{2}}{(\omega +1)a^{4}}\left( \frac{2h}{H}+3\omega -1\right)
, &  \\
\\
\rho =-\frac{b^{2}}{4\pi G(\omega +1)a^{4}}\left( \frac{h}{H}-2\right) , &
\end{array}%
\end{eqnarray}%
where $h=\dot{b}/b$. On the other hand, the conservation of energy-momentum
tensor on the original brane gives
\begin{eqnarray}\label{62}
\rho =\rho _{0}a^{-3(\omega +1)},
\end{eqnarray}%
were $\rho_0$ is energy density in the corresponding epoch. Hence, using equation (\ref{61}) and (\ref{62}) we obtain
\begin{eqnarray}\label{63}
\frac{h}{H}=2-\frac{4\pi G\rho _{0}(\omega +1)}{b^{2}a^{3\omega -1}}.
\end{eqnarray}%
Consequently, using (\ref{61}) and (\ref{63}) we are left with
\begin{eqnarray}\label{64}
b^{2}=\frac{\Lambda }{3}a^{4}+\frac{8\pi G\rho _{0}}{3}a^{1-3\omega }.
\end{eqnarray}%
Also we have
\begin{eqnarray}\label{65}
\frac{1}{R}={\bar{K}}_{\mu \nu }{\bar{u}}^{\mu }{\bar{u}}^{\nu }=\left( 1-%
\frac{h}{H}\right) \frac{b}{a^{2}}.
\end{eqnarray}%
Hence according to (\ref{61}), (\ref{64}) and (\ref{65}) we obtain
\begin{eqnarray}\label{66}
\frac{1}{R}=\frac{4\pi G(3\omega +1)a^{-3(1+\omega )}-\Lambda }{\sqrt{3}%
(8\pi G\rho _{0}a^{-3(1+\omega )}+\Lambda )^{\frac{1}{2}}}.
\end{eqnarray}%
Note that the existence of cosmological constant in the open universe models
$(k=0,-1)$ is necessary  to stabilize the test particles. Inserting equation
(\ref{66}) into   (\ref{56}) shows that in open universe in the absence
of cosmological constant normal curvature in late time universe tends to
the zero and consequently test particles become unstable. According to the resent observations, we live in a flat universe $k=0$. Hence the induced Freedman equation in the radiation dominated universe is
\begin{eqnarray}\label{67}
\dot{a}^{2}=\frac{\Lambda }{3}a^{2}+\frac{8\pi G\rho _{0}}{3a^{2}},
\end{eqnarray}%
with solution
\begin{eqnarray}\label{68}
a^{2}=\sqrt{\frac{8\pi G\rho _{0\gamma }}{\Lambda }}\sinh \left( 2\sqrt{%
\frac{\Lambda }{3}}t\right) .
\end{eqnarray}%
Consequently the normal curvature in the radiation dominated epoch becomes
\begin{eqnarray}\label{69}
\frac{1}{R}=\sqrt{\frac{\Lambda }{3}}\frac{\left( 1-\sinh ^{2}(2\sqrt{\frac{%
\Lambda }{3}}t)\right) }{\sinh (2\sqrt{\frac{\Lambda }{3}}t)\cosh (2\sqrt{%
\frac{\Lambda }{3}}t)}.
\end{eqnarray}%
This equation in the nucleosynthesis epoch $(2\sqrt{\Lambda /3}t\ll 1)$ take
the form
\begin{eqnarray}\label{70}
\frac{1}{R}=\frac{1}{2t}.
\end{eqnarray}%
On the other hand, in the dust dominated universe we obtain from Freedman
equation induced on the original brane the following solution
\begin{eqnarray}\label{71}
a=\left( \frac{8\pi G\rho _{0m}}{\Lambda }\right) ^{\frac{1}{3}}\sinh ^{\frac{2}{3}}\left( \frac{\sqrt{3\Lambda }}{2}t\right) .
\end{eqnarray}%
In this case the normal curvature becomes
\begin{eqnarray}\label{72}
\frac{1}{R}=2\sqrt{\frac{\Lambda }{3}}\frac{1-\sinh ^{2}(\frac{\sqrt{%
3\Lambda }}{2}t)}{\sinh (\sqrt{3\Lambda }t)}.
\end{eqnarray}%
Now equation (\ref{55}) in the nucleosynthesis epoch using approximation takes the
form
\begin{eqnarray}\label{73}
\frac{d^{2}x^4 }{dt^{2}}+\frac{3}{4}t^{-2}x^4 +\frac{1}{2}t^{-1}=0,
\end{eqnarray}%
which have the following exact solution
\begin{eqnarray}\label{74}
x^4 _{\gamma }=\xi _{0\gamma }\sqrt{\frac{t}{t_{1}}}\sin \left( \frac{\sqrt{2%
}}{2}\ln \frac{t}{t_{2}}\right) -\frac{2}{3}t.
\end{eqnarray}%
Here $t_{1}$ and $t_{2}$ are two constants. If we assume $t_{2}=t_{n}$ so
that $t_{n}$ is the nucleosynthesis epoch then we have
\begin{eqnarray}\label{75}
u_{\gamma }^{4}=\frac{dx^4 }{dt}\vert _{t=t_{n}}=\frac{\sqrt{3}\xi _{0\gamma }%
}{2\sqrt{tt_{1}}}\sin \left( \frac{\sqrt{2}}{2}\ln \left( \frac{t}{t_{n}}%
\right) +\theta \right) -\frac{2}{3},
\end{eqnarray}%
where $\tan \theta =\sqrt{2}$. In the above relation $\xi _{0\gamma }$
denotes the width of the brane in the nucleosynthesis duration. Hence
\begin{eqnarray}\label{76}
u_{\gamma }^{4}(t_{n})\sim -\frac{2}{3}.
\end{eqnarray}%
Also the corresponding solution of equation (\ref{55}) in the present epoch becomes
\begin{eqnarray}\label{77}
u_{m}^{4}(t_{0})\sim -\frac{1}{3}\frac{dR}{dt}\vert_{t=t_{0}}=-\frac{1}{2}
\frac{\sinh ^{2}(\frac{\sqrt{3\Lambda }}{2}t)}{[1-\sinh ^{2}(\frac{\sqrt{%
3\Lambda }}{2}t)]^{2}}\vert_{t=t_{0}}.
\end{eqnarray}%
The age $t_{0}$ of the universe can be found by the condition $a(t_{0})=1$.
Using the identity $\tanh ^{-1}x=\sinh ^{-1}(\frac{x}{\sqrt{1-x^{2}}})$ we
get the expression
\begin{eqnarray}\label{78}
t_{0}=\frac{2}{\sqrt{3\Lambda }}\tanh ^{-1}\sqrt{\Omega _{\Lambda }}.
\end{eqnarray}%
Inserting the values $t_{0}=13.7\times 10^{9}$ years and $\Omega _{\Lambda
}=0.7$ found from the WMAP measurements of the temperature fluctuations in
the cosmic microwave background radiation, and from the determination of the luminosity-redshift relationship of supernova of type Ia, we get $\frac{%
8\pi G_{N}}{3}\rho _{0}=\frac{1-\Omega _{\Lambda }}{\Omega _{\Lambda }}=0.43$
and $\Lambda =1.1\times 10^{-20}$ years$^{-2}$. Consequently the velocity of particles in the present epoch along the extra dimension becomes
\begin{eqnarray}\label{79}
u_{m}^{4}(t_{0})\sim -0.662.
\end{eqnarray}%
We can now estimate the variation of mass from nucleosynthesis up to the
present epoch. Defining the quotient $\frac{\Delta m}{m}$ as
\begin{eqnarray}\label{80}
\frac{\Delta m}{m_0}=\frac{m(t_{0})-m(t_{n})}{m(t_{0})},
\end{eqnarray}%
where $t_{0}$ and $t_{n}$ denote the age of universe and time of
nucleosynthesis respectively. Now using equation (\ref{51}),  (\ref{76}) and (\ref{79}) we have
\begin{eqnarray}\label{81}
\frac{\Delta m}{m_0}\sim 1-e^{0.004}\sim -0.004.
\end{eqnarray}%
If we consider that a variation of the rest mass of  particles had occurred
between the epoch of primordial nucleosynthesis and present, then one can
calculate the deviation in the $^4He$ production from the Hot Big Bang model
prediction with this fact. According to  \cite{10} if we the masses are
changed then the upper bound of the deviation of primordial Helium production due to a variation on the difference between the rest mass of the nucleons between the present and nucleosynthesis epoches is given by 
\begin{eqnarray}\label{82}
\delta (M_{n}-M_{p})\leq 0.129MeV,
\end{eqnarray}%
where $M_{n}$ and $M_{p}$ are neutron and proton masses respectively. If we define $\Delta Q=M_{n}-M_{p}$, then the above limit gives \cite{10}
\begin{eqnarray}\label{83}
\left\vert \frac{\delta (\Delta Q)}{\Delta Q_{0}}\right\vert \leq 10\%,
\end{eqnarray}%
with
\begin{eqnarray}\label{83a}
\Delta Q_0\simeq 294 Mev.
\end{eqnarray}
The authors of \cite{10} used the $5D$ induced matter brane cosmological
model with compact extra dimension in the radiation dominated universe. Then
using the original definition of induced mass (\ref{5}) we find the following
quotient for the mass variation of nucleons 
\begin{eqnarray}\label{83b}
\left\vert \frac{\delta (\Delta Q)}{\Delta Q_{0}}\right\vert \approx 100\%,
\end{eqnarray}
which is in disagreement with the previous bound (\ref{83}).
Note that, if we use  equation (\ref{42}) or equivalently (\ref{51})
as variation of mass, then in compact models of IMT \cite{27} the mass of particles remain unchanged. On the other hand, in noncompact IMT, equation (\ref{81}) gives a better outcome
\begin{eqnarray}\label{84}
\left\vert \frac{\delta (\Delta Q)}{\Delta Q_{0}}\right\vert \approx 0.4\%,
\end{eqnarray}%
which is in agreement with Hot Big Bang result (\ref{83}). 
\section{Conclusions}
In this paper we have analysed the variation of nucleon masses in flat FRW
cosmological model, embedded in a $5D$ Ricci flat bulk space,  using IMT
ideas. We have showed that the mass variation is a consequence of non-integrability
of the $4D$ embedded spacetime. From the point of view of a $4D$ observer, according to the (\ref{42}), the  variation of the mass of particles is a direct result of the distribution of matter in $4D$ universe. This relation can be regarded as an explanation of mach''s principle, that inertial forces should be generated by the motion of a body relative to the bulk of induced matter in the universe. In the
theory outline in this paper, the variation of mass obtained from IMT is in agreement with mass variation bound obtained from Hot Big Bang.

\end{document}